\documentclass{elsart}
\usepackage{psfig}
\DeclareRobustCommand{\element}{\relax\ifmmode\@tempswafalse
\else\@tempswatrue\fi\clearelargs\def\?{\phantom{0}}\@lement}
\def\@lement#1{\if#1[\expandafter\f@@dargs\else\druck@lement{#1}\fi}

\DeclareMathAlphabet{\mathsc}{OT1}{cmr}{m}{sc}
\def\testbx{bx}%
\DeclareRobustCommand{\ion}[2]{%
\relax\ifmmode
\ifx\testbx\f@series
{\mathbf{#1\,\mathsc{#2}}}\else
{\mathrm{#1\,\mathsc{#2}}}\fi
\else\textup{#1\,{\mdseries\textsc{#2}}}%
\fi}

\begin{document}

\begin{frontmatter}
\title{Modelling of the X-ray broad absorption features in Narrow Line
 Seyfert 1s}

\author{Delphine Porquet}
\address{Service d'Astrophysique, CEA, Saclay, France}
\author{Martine Mouchet, Anne-Marie Dumont}
\address{DAEC, Observatoire de Paris, section de Meudon, France}
\begin{abstract}
We investigate the origin of the broad absorption features detected 
near 1-1.4 keV in several Narrow Line Seyfert 1 galaxies, by modelling 
the absorbing medium with various physical parameters, using the ionisation 
code PEGAS. The observed properties of the X-ray absorption features can be  
reproduced by taking into account the peculiar soft X-ray excess which is 
well fitted by a blackbody plus an underlying power law. We equally 
stress that the emission coming from the absorbing medium 
(related to the covering factor) has a strong influence on the 
resulting X-ray spectrum, in 
particular on the apparent position and depth of the absorption 
features. A non-solar iron abundance may be required to 
explain the observed deep absorption. We also investigate the
 influence of an additional collisional ionization process (``hybrid case'') on the predicted absorption features.\\
\end{abstract}
\end{frontmatter}

\section{Introduction}
A systematic analysis of the X-ray spectral properties of a sample of 22
so-called Narrow Line Seyfert\,1s (NLS1s) based on ASCA observations has
shown evidence for a broad absorption feature centred in the energy range
1.1-1.4\,keV, in 6 of the 22 objects \cite{Vaughan99}. The absorption
feature strength is typically about 100\,eV with an intrinsic width
ranging from 0.1 to 0.3\,keV. Until now, this type of absorption has
never been detected in Broad Line Seyfert\,1s (BLS1s) spectra. Only 3 NLS1s
exhibit absorption edges in the 0.7-0.9\,keV range consistent with that 
seen in at least 50\% of the BLS1s. Several explanations of the
1.1-1.4\,keV absorption features have been proposed: a blueshift of
\ion{O}{vii}--\ion{O}{viii} edges or lines (outflow: z$\sim$0.2-0.6\,c)
\cite{Leighly97,Ulrich99}; resonance absorption lines from Mg, Si, S and
Fe\,L \cite{Nicastro99b,Turner99}; an enhancement of Fe or Ne
\cite{Leighly97,Ulrich99}.

\section{The models and results}
\noindent The calculations are made with the photoionization code PEGAS adapted for optically thin media
\cite{DumontPorquet98}.\\
A photoionized plasma is characterized by its ionization parameter:
$\xi$=$\frac{L}{n_{H}~R^{2}}$, where L is the bolometric incident
luminosity (erg\,s$^{-1}$), n$_{\mathrm{H}}$ (cm$^{-3}$) is the hydrogen
density and R (cm) is the distance of the inner layer of the cloud from
the ionizing incident source.\\

\noindent $\bullet$ \underline{Influence of the incident continuum shape}\\
\noindent NLS1s tend to be stronger soft X-ray emitters with respect to
BLS1, and
often have steeper photon indices: $\Gamma$(2-10\,keV)=1.6--2.5
\cite{Vaughan99}. Their soft excesses can be modelled as blackbody (BB)
emission superposed on an underlying power law (PL). We consider a
``typical incident NLS1 continuum'' as the sum of a BB at T=130\,eV and
a PL with $\alpha=1$ (F(E)\,=\,E$^{-\alpha}$) and with L$_{\rm
BB}/L_{\rm PL}$(2-10 keV)=3. For a given $\xi$ and N$_{H}$ (column
density in cm$^{-2}$) the ``typical incident NLS1 continuum'' gives less
absorption and higher ionization states than a single PL with the same
slope. This could explain why the absorption features in NLS1s are located at
higher energies.

\noindent $\bullet$ \underline{Influence of iron overabundance}\\
\begin{figure}
\begin{center}
\hspace{0.9cm}\psfig{file=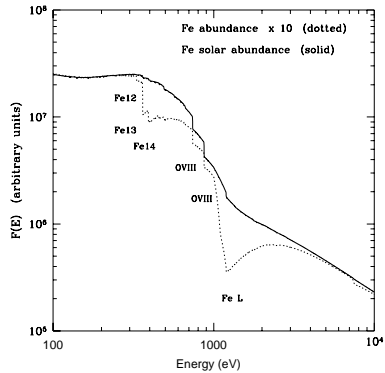,height=7cm}
\end{center}
\caption{Transmitted spectra corresponding to a typical
NLS1 incident continuum obtained with a solar iron abundance (solid curve)
and with an iron overabundance of a factor 10 (dotted curve). $\xi$=50 ,
N$_{H}$=10$^{22}$ cm$^{-2}$,
and E/$\Delta$E=300.}
\label{fig2}
\end{figure}
Figure~1 shows the transmitted spectrum obtained for an iron
overabundance of a factor 10, as well as the spectrum for a solar iron
abundance. An iron overabundance could be responsible for a strong
absorption above 1\,keV (\ion{Fe}{xvii} + some \ion{Ne}{ix}) and a weak
oxygen absorption.

\noindent $\bullet$ \underline{Influence of emission (covering factor)}\\
The emission ($\varepsilon$) depends on the covering factor
(f$_{{\mathrm{c}}}$) of the medium responsible for the absorption
($\varepsilon\propto$~f$_{\mathrm{c}}$). Figure~2 illustrates the great
importance of the emission lines on the observed spectrum.\\

\noindent $\bullet$ \underline{Influence of collisional processes}\\
\noindent The Warm Absorber in BLS1s could be purely
 photoionized, but an additional ionization process such as collision or a
 non pure radiative equilibrium are not ruled out
 \cite{PorquetDumont98,Porquet99,Nicastro99a}. This could be also the
 case in NLS1s. As shown in Figure~\ref{fig2}, a pure photoionized case
 and a hybrid case (T\,=\,3.6~$10^{6}$K) have different line ratios and
 different profiles of absorption (for the same incident continuum and
 $\xi$). In the hybrid case, higher ionisation states occur (i.e.
 \ion{Fe}{xvii}, \ion{Ne}{x}) and for a non negligible covering factor (f
 $>$ 0.5), the spectrum exhibits an absorption above 1\,keV but no oxygen
 edges, which could explain the lack of detection of oxygen edges in the
 spectra of the 6 objects with broad absorption above 1\,keV.

\begin{figure}
\hspace{1.5cm}\psfig{file=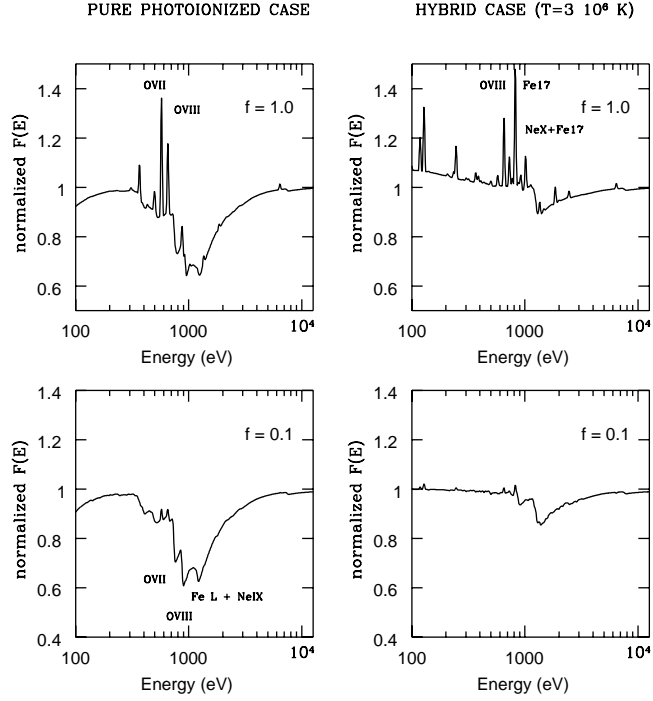,height=10cm}
\caption{Spectra normalized to the incident continuum,
for a pure photoionized medium (left) and for a hybrid case with T fixed
at 3.6~10$^{6}$\,K (right). Bottom: a covering factor f=0.1;
top: f=1.0. Model parameters are $\xi$=50, N$_{H}$=10$^{22}$cm$^{-2}$,
and $E/ \Delta E$=30.}
\label{fig2}
\end{figure}

\section{Conclusion and perspectives}

Several properties of NLS1s eg. UV-X-ray energy distribution, iron
overabundance, hybrid plasma with high covering factor, could account for
the peculiarities observed in some soft X-ray spectra of these objects.
The spectral resolution attainable by ASCA was insufficient to disentangle
between these different possibilities. The new X-ray satellites (Chandra,
XMM-Newton) offer the prospect of detailed spectra that will certainly allow
us to determine the nature of the 1\,keV feature (emission or
absorption). Moreover X-ray spectroscopic diagnostics such as those based
on the ratios of He-like ion lines will enable us to determine the
ionizing process (either pure photoionization, or photoionization plus an
additional ionization process), as well as the gas density
\cite{PorquetDubau2000b,Kaastra2000}. The determination of the physical
parameters for the Warm Absorber media in NLS1s and in BLS1s will provide
constraints on unified schemes.\\

\end{document}